\documentclass[aps,prl,twocolumn,superscriptaddress,showpacs,preprintnumbers,amsmath,amssymb,floatfix]{revtex4-1}

\usepackage{graphicx} % Include figure files
\usepackage{dcolumn}  % Align table columns on decimal point
\usepackage{subfigure}
\usepackage{chngcntr}
\usepackage{makecell}
\usepackage{color}
\usepackage{CJK}
\usepackage{indentfirst}
\usepackage{amsmath}
\usepackage{float}
\usepackage{lineno}
\usepackage{verbatim}
\usepackage{orcidlink}
\usepackage{xcolor}
\usepackage{ulem}

\graphicspath{{ps}}

\begin{document}

%\vspace*{-3\baselineskip}
%\resizebox{!}{3cm}{\includegraphics{belle.eps}}

\title{ \quad\\[1.0cm] Evidence for $B^0 \to p\bar{\Sigma}^0\pi^-$ at Belle}

%%%% >>>>> insert the authorlist here. BEFORE the abstract !!!!! <<<<<
%%%% >>>>> from the authorship confirmation web page
%%% Name the file author.tex and use \input{author} to insert into your latex file.
%\author{author}
%\input{pub648-orcid}
%\affiliation{affiliation}
%\collaboration{The Belle Collaboration}
\noaffiliation
  \author{C.-Y.~Chang\,\orcidlink{0000-0003-1614-1708}} % 11983
  \author{M.-Z.~Wang\,\orcidlink{0000-0002-0979-8341}} % 2074
  \author{I.~Adachi\,\orcidlink{0000-0003-2287-0173}} % 2590
% \author{K.~Adamczyk\,\orcidlink{0000-0001-6208-0876}} % 2239
% \author{J.~K.~Ahn\,\orcidlink{0000-0002-5795-2243}} % 7423
  \author{H.~Aihara\,\orcidlink{0000-0002-1907-5964}} % 2223
  \author{S.~Al~Said\,\orcidlink{0000-0002-4895-3869}} % 6823
  \author{D.~M.~Asner\,\orcidlink{0000-0002-1586-5790}} % 4684
  \author{H.~Atmacan\,\orcidlink{0000-0003-2435-501X}} % 2538
% \author{V.~Aulchenko\,\orcidlink{0000-0002-5394-4406}} % 8183
  \author{T.~Aushev\,\orcidlink{0000-0002-6347-7055}} % 3747
  \author{R.~Ayad\,\orcidlink{0000-0003-3466-9290}} % 3766
% \author{T.~Aziz\,\orcidlink{-}} % 3523
  \author{V.~Babu\,\orcidlink{0000-0003-0419-6912}} % 5623
% \author{S.~Bahinipati\,\orcidlink{0000-0002-3744-5332}} % 2332
% \author{A.~M.~Bakich\,\orcidlink{0000-0001-8315-4854}} % 2115
% \author{Y.~Ban\,\orcidlink{-}} % 3503
% ...
% 2387
% \author{V.~Zhulanov\,\orcidlink{0000-0002-0306-9199}} % 4983
\collaboration{The Belle Collaboration}

%% end author list

%\linenumbers

\begin{abstract}
 We search for the $B^0\to p\bar{\Sigma}^0\pi^-$ decay with $\bar{\Sigma}^0 \to
\bar{\Lambda}\gamma$, where the $\gamma$ is not measured, using a data sample corresponding to an integrated luminosity of 711 $\rm{fb^{-1}}$ which contains 772 $\times$ $10^{6}$ $B\bar{B}$ pairs, collected around the $\Upsilon$(4S) resonance with the Belle detector at the KEKB asymmetric-energy $e^{+}e^{-}$ collider. We measure for the first time the $B^0\to p\bar{\Sigma}^0\pi^-$ branching fraction to be $\mathcal{B}(B^0 \to p \bar\Sigma^0  \pi^-) = (1.17^{+0.43}_{-0.40}(\text{stat})\pm 0.07(\text{syst})) \times 10^{-6}$ with a significance of $3.0\sigma$. We simultaneously measure the branching fraction for the related channel $B^{0}\to p\bar{\Lambda}\pi^{-}$ with much improved precision. 

\end{abstract}

%\pacs{XX.YY.ZZ, AA.BB.CC}

\maketitle

%%%% >>>> keep the final version single-spaced
\tighten

{\renewcommand{\thefootnote}{\fnsymbol{footnote}}}
\setcounter{footnote}{0}

% {\it SVD1:}
% {The Belle detector is a large-solid-angle magnetic
% {spectrometer that
% {consists of a three-layer silicon vertex detector (SVD),
% {a 50-layer central drift chamber (CDC), an array of
% {aerogel threshold Cherenkov counters (ACC), % <- \v{C}erenkov 2007.08
% {a barrel-like arrangement of time-of-flight
% {scintillation counters (TOF), and an electromagnetic calorimeter
% {comprised of CsI(Tl) crystals (ECL) located inside 
% {a super-conducting solenoid coil that provides a 1.5~T
% {magnetic field.  An iron flux-return located outside of
% {the coil is instrumented to detect $K_L^0$ mesons and to identify
% {muons (KLM).  The detector
% {is described in detail elsewhere~\cite{Belle}.

% {\it SVD2+SVD1:}
%
%%%%%%%%%%%%%%%%%%%%%%%%%%%%%%%%%%%%%%%%%%%%%%%%%%%%%%%%%%%%%%%%%%5
%
%  NOTE: The Publication Council affirms (14/6/2021) that
%  this text is in the PRESENT TENSE.
%
%%%%%%%%%%%%%%%%%%%%%%%%%%%%%%%%%%%%%%%%%%%%%%%%%%%%%%%%%%%%%%%%%%5
%

\begin{figure}[ht]
    \centering
    \includegraphics[width =6cm]{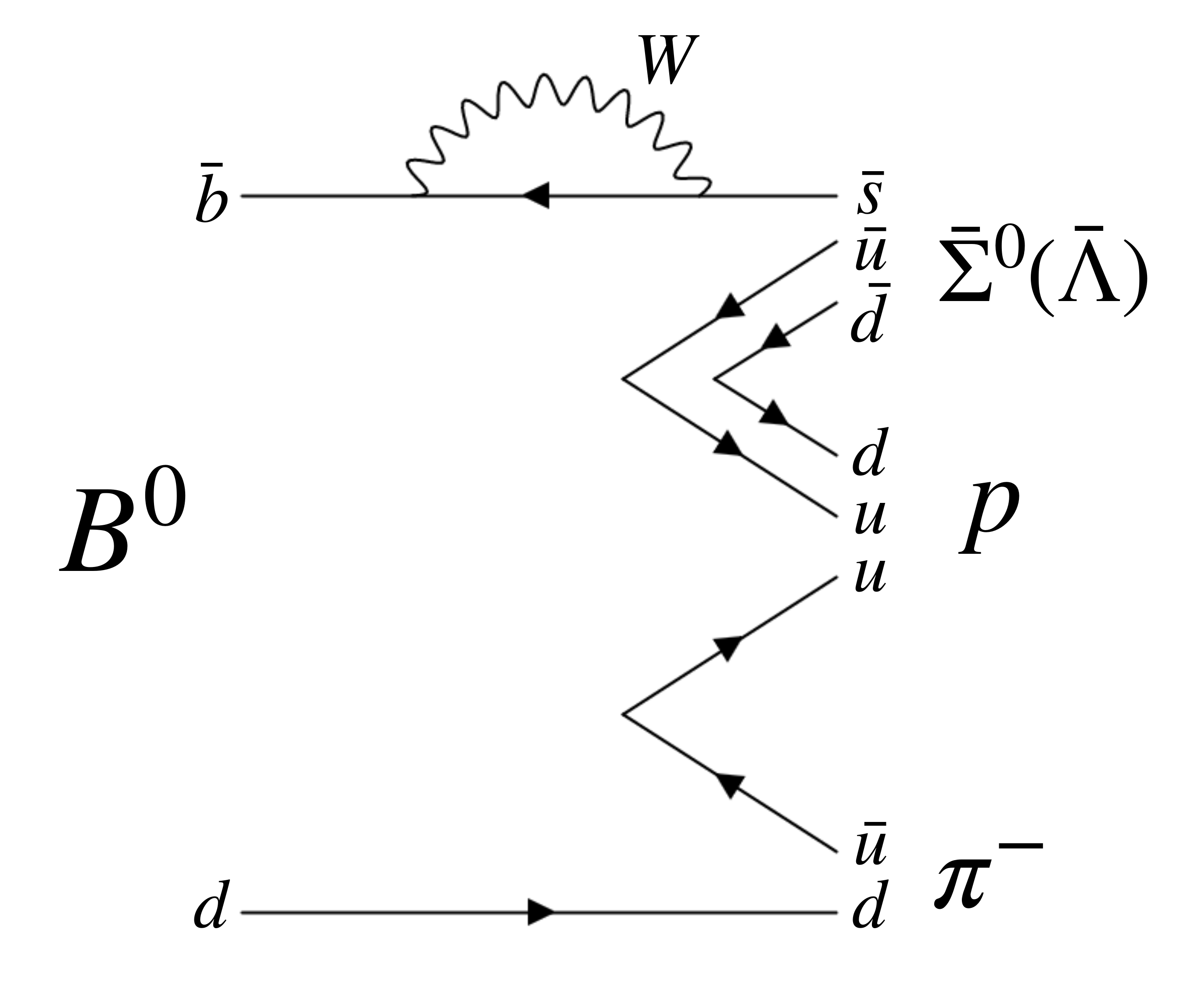}
    \caption{A possible $b\to s g$ penguin diagram for $B^0\to p\bar{\Sigma}^0\pi^-$ and $B^0\to p\bar{\Lambda}\pi^-$.}
    \label{fig:my_label}
\end{figure}

Since the first observation of the rare baryonic $B$ decay $B^{+}\to p\bar{p}K^{+}$ ~\cite{PhysRevLett.88.181803}, many other such decay modes, together with their puzzling features, have been found which are summarized in Ref.~\cite{Legacy} section 17.12. One unexpected feature is the observation of an asymmetry in the angular distribution in $B^0 \to  p \bar\Lambda  \pi^-$~\cite{PhysRevD.76.052004}. This indicates that the proton tends to move faster than the accompanying $\bar\Lambda$ in the $B$ meson rest frame, which contradicts the intuitive $b \to s g$ picture~\cite{annurev.nucl.48.1.253} of the $B^0 \to  p \bar\Lambda  \pi^-$ decay which is shown in Fig \ref{fig:my_label}. Factorization approaches and QCD counting rules~\cite{PhysRevD.66.014020,PhysRevD.66.054004} predict that the branching fraction ($\mathcal{B}$) of $B^0 \to  p \bar\Lambda  \pi^-$ is much smaller than that of $B^0 \to p \bar\Sigma^0  \pi^-$, and is an order of magnitude smaller than the current measured branching fraction of $(3.23^{+0.33}_{-0.29}\pm 0.29)\times 10^{-6}$~\cite{PhysRevD.76.052004}. A similar approach with modifying the axial-vector and pseudo-scalar operators~\cite{Hou}, predicts $\mathcal{B}( B^0 \to  p \bar\Lambda  \pi^- )$ and $\mathcal{B}(B^0 \to p \bar\Sigma^0  \pi^-) \sim 1.6 \times 10^{-6}$. Thus, it is promising to observe $\mathcal{B}(B^0 \to p \bar\Sigma^0  \pi^-)$ using the full $\Upsilon(4S)$ data set collected by the Belle experiment. Another important aspect for this study is the large $B$ sample accumulated by the LHCb experiment, where branching fraction measurements of rare baryonic $B$ decays such as $B^{0}_{s} \to p\bar{\Lambda}K^- $~\cite{PhysRevLett.119.041802} utilize $\mathcal{B}(B^0 \to  p \bar\Lambda  \pi^-)$ for the normalization. These measurements will hence profit from a more precise measurement of $\mathcal{B}(B^0 \to  p \bar\Lambda  \pi^-)$. In addition, the systematic uncertainty due to the contamination of $B^0 \to p \bar\Sigma^0  \pi^-$  can be better constrained if $\mathcal{B}(B^0 \to p \bar\Sigma^0  \pi^-)$ is known.\par
We study the decay $B^0 \to p\bar{\Sigma}^0\pi^-$ based on a data sample that corresponds to an integrated luminosity of 711$\ \rm{fb^{-1}}$ and contains $772 \times 10^{6}$ $B\bar{B}$ pairs. The whole data sample is collected with the Belle detector at the KEKB asymmetric-energy $e^{+}e^{-}$ collider \cite{KUROKAWA20031,ptep} using an $e^+$ energy of 3.5 GeV and an $e^-$ energy of 8 GeV. The Belle detector~\cite{Belle:2000cnh,belle} is a large-solid-angle magnetic spectrometer which includes a silicon vertex detector (SVD), a 50-layer central drift chamber (CDC), an array of 44 aerogel threshold Cherenkov counters (ACC), a barrel-like arrangement of time-of-flight scintillation counters (TOF), an electromagnetic calorimeter (ECL) comprised of CsI(Tl) crystals located inside a superconducting solenoid coil that provides a 1.5 T magnetic field, and an iron flux-return located outside of the coil, which is instrumented to detect $K_{L}$ mesons and identify muons (KLM).  The details of each subdetector are described in Ref.~\cite{Belle:2000cnh}. \par 
% monte carlo part 
We use Monte Carlo (MC) samples to study the selection criteria for signal reconstruction. The MC samples are generated by the \texttt{EvtGen} package~\cite{Lange:2001uf} and the response of the Belle detector is simulated by the \texttt{GEANT3} package~\cite{Brun:1987ma}. In signal MC sample, one of the $B^0$ or $\bar{B}^0$ mesons in the $B^0\bar{B}^0$ pair decays to our signal channel, and the other one decays randomly according to the known or estimated branching fractions. We generate signal MC samples using two decay models: a threshold-enhancement (TE) model, where the baryon-antibaryon pair in the final state is generated near its kinematic threshold, and a phase-space (PHSP) model, where the final-state particles from the decay are distributed uniformly in phase space. In the TE model, we define hypothetical particles with nominal mass values of $2.05\ \text{GeV}/c^{2}$ and $2.18\ \text{GeV}/c^{2}$, and a width of $0.3\ \text{GeV}/c^{2}$, which decay to $p\bar{\Lambda}$ and $p\bar{\Sigma}^0$, respectively. In our analysis, we assume the signal events to follow the TE model. The classification of signal and background events is optimized using signal MC samples where we assumed a $B^0\to p\bar{\Sigma}^0\pi^-$ branching fraction of $1.9\times 10^{-6}$, which is half the value of the previously measured upper limit~\cite{Belle:2003taw}. To study the background, we use MC samples that include generic $B$ decays (both $B$ mesons are collection of $b\to c$ process only, with known or estimated branching fractions), continuum background ($e^{+}e^{-} \to q\bar{q}\ $ with $q = u,d,s,c$), and rare $B$ decays ($b\to u,d,s$). \par

% pre-selection 
The charged-track selection criteria are based on the information obtained from the tracking system (SVD and CDC). We use information from the CDC, ACC, and TOF to form hadron particle identification (ID) \cite{pid} likelihood values $\mathcal{L}_i$ calculated for each particle hypothesis~$i$---with $i = \pi, K$, and $p$. For lepton (e,\ $\mu$) particle ID, the information from the KLM is used to calculate muon ID likelihood value $\mathcal{L}_{\mu}$, and the information from the CDC, ACC and ECL is utilized to form electron ID likelihood value $\mathcal{L}_{e}$. We first reject highly electron-like ($\mathcal{L}_{e} \ge $ 0.95) and muon-like ($\mathcal{L}_{\mu} \ge $  0.95) tracks. To select proton-like tracks, we require $\mathcal{L}_p/(\mathcal{L}_{p}+\mathcal{L}_{K}) > 0.6$ and $\mathcal{L}_p/(\mathcal{L}_{p}+\mathcal{L}_{\pi}) > 0.6$ which has $92.4\%$ selection efficiency. To select $\pi$-like tracks, we require $\mathcal{L}_K/(\mathcal{L}_{\pi}+\mathcal{L}_{K}) < 0.4$ which has $92.1\%$ selection efficiency. We constrain the distance of closest approach of a charged track and the interaction point (IP) along the $e^{+}$ beam direction ($dz$) and in the plane transverse to it ($dr$). For protons (pions), which do not originate from a $\Lambda$ decay, we require $|dr| < 0.3 (0.3)$ cm and $|dz| < 2 (4)$ cm.\par

%Lambda selection 
The $\Lambda$ is reconstructed by combining a $p$ track and a   $\pi$ track. We select $\Lambda$ candidates by applying $\Lambda$-momentum-dependent constraints on four kinematic variables: the distance between the two daughter tracks at their interception position along the $e^{+}$ beam direction; the smaller of the proton $dr$ and pion $dr$; the angle between the vector from the primary to the secondary vertex and the momentum vector of the $\Lambda$; and the flight-path length of the $\Lambda$ candidate. The invariant mass of the $\Lambda$ candidates is required to be in the range from 1.111 to 1.121~$\text{GeV}/c^{2}$, which corresponds approximately to the $\pm3\sigma_{m_\Lambda}$ interval with $\sigma_{m_\Lambda}$ being the $\Lambda$ mass resolution. The daughter proton of the $\Lambda$ candidate is required to have $\mathcal{L}_p/(\mathcal{L}_{p}+\mathcal{L}_{K}) > 0.6$ and  $\mathcal{L}_p/(\mathcal{L}_{p}+\mathcal{L}_{\pi}) > 0.6$ in order to reduce combinatorial background. But we don't impose any additional criteria on pions.\par

% Reconstruction 
The $B$ signal candidates are reconstructed by combining the 4-momenta of $p$, $\bar\Lambda$ and $\pi^{-}$ candidates including their charged-conjugate states. Soft photon candidates from $\bar{\Sigma}^0 \to \bar{\Lambda}\gamma$ are not included in the signal $B$ reconstruction, since this would lead to excessive background. For each $B$ candidate, a kinematic fit is performed, which constrains the 4-momenta of all final-state particles to originate from a common vertex. Only $B$ candidates where this fit converges are selected.\par 

We also measure $B^0 \to p \bar{\Lambda} \pi^-$ simultaneously to obtain a more precise branching fraction and validate the analysis procedure for $B^0 \to p\bar{\Sigma}^0 \pi^-$. To disentangle the $B^0\to p\bar{\Sigma}^0\pi^-$ contribution from the $B^0 \to  p \bar\Lambda  \pi^-$ contribution and the background contributions, we use the distribution of two kinematic variables defined in the $e^+e^-$ center-of-momentum (c.m.) frame: the beam-energy-constrained mass $M_\text{bc}$ $= \sqrt{E^{2}_{\rm beam}/c^{4}-p^{2}_{B}/c^{2}}$, and the energy difference $\Delta E$ $= E_{\rm beam}-E_{B}$, where $E_{\rm beam}$ is the beam energy and  $p_{B}$, $E_{B}$ are the momentum and the energy of the $B$ meson candidate, respectively. We require our signal candidates to lie within $\Delta E\in (-0.14,0.20)\ \text{GeV}$, which effectively reduces the contributions from possible rare channels and other excited states like $B^0\to p\bar{\Lambda}(1405)\pi^-$ and $B^0 \to p\bar{\Sigma}^0(1385)\pi^-$, such that they are insignificant compared to signals in MC simulation, and $M_\text{bc}\in (5.23,5.30)\ \text{GeV}/c^{2}$.\par

%background study 
Using the background MC samples described above, we find that the dominant source of background comes from $q\bar{q}$ continuum background. In contrast to $b\to c$ and $b\to u,d,s$ decays of $B$ mesons, continuum background can be distinguished well from signal events by its shape variables. To classify signal and background events, we use a multivariate analysis package named NeuroBayes \cite{FEINDT2006190}, which is based on neural networks. The training samples are generated using a signal MC sample and a background sample obtained from real data in a sideband region defined by $\Delta E>0.1\ \rm{GeV}$. The neural network was trained using the following 29 input variables: 20 modified Fox-Wolfram moments \cite{PhysRevLett.41.1581,belle_sfw}; the cosine of polar angle of the $B$ daughters with respect to the beam axis; the cosine of the angle between the reconstructed $B$ flight direction in the beam axis; the sphericity of the event calculated in the c.m. frame \cite{PhysRevD.1.1416}; the longitudinal distance between the vertices of signal $B$ candidate and the accompanying $\bar{B}$; the quality variable given by the flavor tagging used to judge how likely a $B$ candidate is a $B $ meson\cite{Belle:2004uxp}; as well as the missing mass and missing energy of the event obtained from the deviation between the reconstructed and expected four momentum of the accompanying $\bar{B}$. The output discriminant of NeuroBayes varies from $-1$ to $+1$, where a value close to +1 corresponds to a signal-like event, whereas $-1$ corresponds to a background-like event. We determine the threshold on the NeuroBayes output by optimizing the figure-of-merit (FOM) given as 

\begin{equation}
     \text{FOM} = \frac{N_{\rm{sig}}}{\sqrt{ N_{\rm{sig}}+N_{\rm{bkg}}}} ,
\end{equation} 
where $N_{\rm sig}$ is the number of signal events estimated from the signal efficiency and assumed branching fraction  in the MC simulation, and $N_{\rm bkg}$ is the number of background events obtained from the $q\bar{q}$ continuum background MC scaled to the full Belle luminosity in the signal region $\Delta E\in (-0.14,0.00)\ \text{GeV}$ and $M_\text{bc}\in (5.26,5.30)\ \text{GeV}/c^{2}$ and calibrated by a scaling from $\Delta E$ sidebands in data into $\Delta E$ signal region. \par
To remove the background from $B^0 \to p \Lambda^{-}_c$ without a major loss of reconstruction efficiency, we veto events within $M_{\Lambda\pi} \in (2.15,2.30)\  \text{GeV}/c^2$, which corresponds approximately to the $\pm2\sigma_{m_{\Lambda\pi}}$ interval around the $\Lambda^{-}_c$ nominal mass with $\sigma_{m_{\Lambda\pi}}$ being the $\Lambda^{-}_c$ mass resolution. \par
For events with multiple signal candidates, we select the candidate with the smallest vertex $\chi^{2}$ from the $B$ vertex fit by using $p$, $\pi$ tracks and $\Lambda$ vertex reconstructed with $p$,$\pi$ tracks of $\Lambda$ candidates. MC studies show that this selects the true candidate in 99.3$\%$ of the events with more than one candidate. In the real data, $3.66\%$ of events have multiple $B$ candidates with an average multiplicity of 1.04.\par

To extract the signal yields, we perform a two-dimensional (2D) extended unbinned maximum likelihood fit to the $(\Delta E, M_\text{bc})$ distribution. The likelihood function is given by  
\begin{equation} \label{eq:2}
\begin{aligned}
\mathcal{L} = \frac{e^{-{(N_{s1}+N_{s2}+N_{b})}}}{N!}\prod^{N}_{i=1}[N_{s1}P_{s1}(M^{i}_\text{bc},\Delta E^{i}) \\
+N_{s2}P_{s2}(M^{i}_\text{bc},\Delta E^{i})+N_{b}P_{b}(M^{i}_\text{bc},\Delta E^{i}) \\
+N_{cb}P_{cb}(M^{i}_\text{bc},\Delta E^{i})],
\end{aligned}
\end{equation}
where $i$ denotes the $i^{\rm th}$ event and $N$ is number of measured events. $N_{s1}$, $N_{s2}$, $N_{b}$ and $N_{cb}$ are the yields for $B^0\to p\bar{\Sigma}^0\pi^-$, $B^0\to p\bar{\Lambda}\pi^-$, background and combinatorial background of $B^0 \to p\bar{\Lambda}\pi^-$, respectively. $P_{s1}$, $P_{s2}$, $P_{b}$ and $P_{cb}$ represent the probability density functions (PDF) for $B^0\to p\bar{\Sigma}^0\pi^-$, $B^0\to p\bar{\Lambda}\pi^-$, background and combinatorial background of $B^0 \to p\bar{\Lambda}\pi^-$, respectively. \par
The following PDF shapes are obtained from MC simulation. We model $P_{s1}$ using a 2D binned PDF obtained from MC simulation. For the $B^0 \to p \bar{\Lambda} \pi^-$ channel, we use a double Gaussian with a common mean to model the $M_\text{bc}$ distribution and a triple Gaussian  with a common mean to model the $\Delta E$ distribution. Because the $\Delta E$ and $ M_\text{bc}$ distributions of the $q\bar{q}$ continuum background and the other $B$ backgrounds are similar, we combine these components and use an ARGUS function~\cite{ARGUS:1990hfq} to describe $M_\text{bc}$ and a second-order polynomial to describe $\Delta E$. To model the combinatorial background caused by $B^0 \to p\bar{\Lambda}\pi^-$, we use a 2D nonparametric PDF using Kernel Estimation~\cite{Cranmer:2000du} obtained from MC simulation. The free parameters in the fit are yields of the 4 components and the shape parameter of ARGUS function.\par  

\begin{figure}[htb]
\centering
\subfigure[]
{
\begin{minipage}[c]{0.25\textwidth}%调节数值可以改变两图之间的距离
\centering
\includegraphics[width=4.5cm]{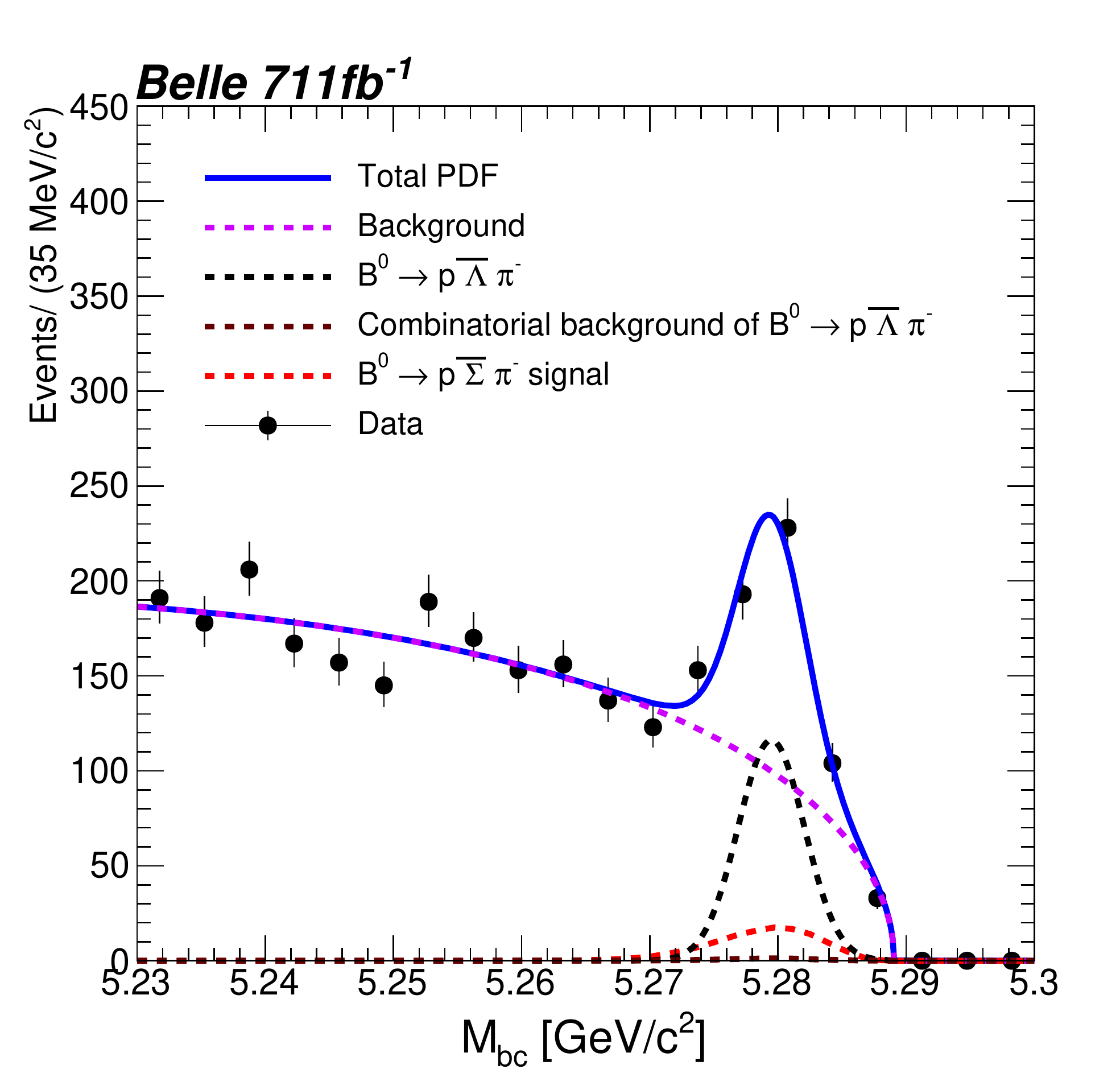}
\end{minipage}%
\begin{minipage}[c]{0.25\textwidth}
\centering
\includegraphics[width=4.5cm]{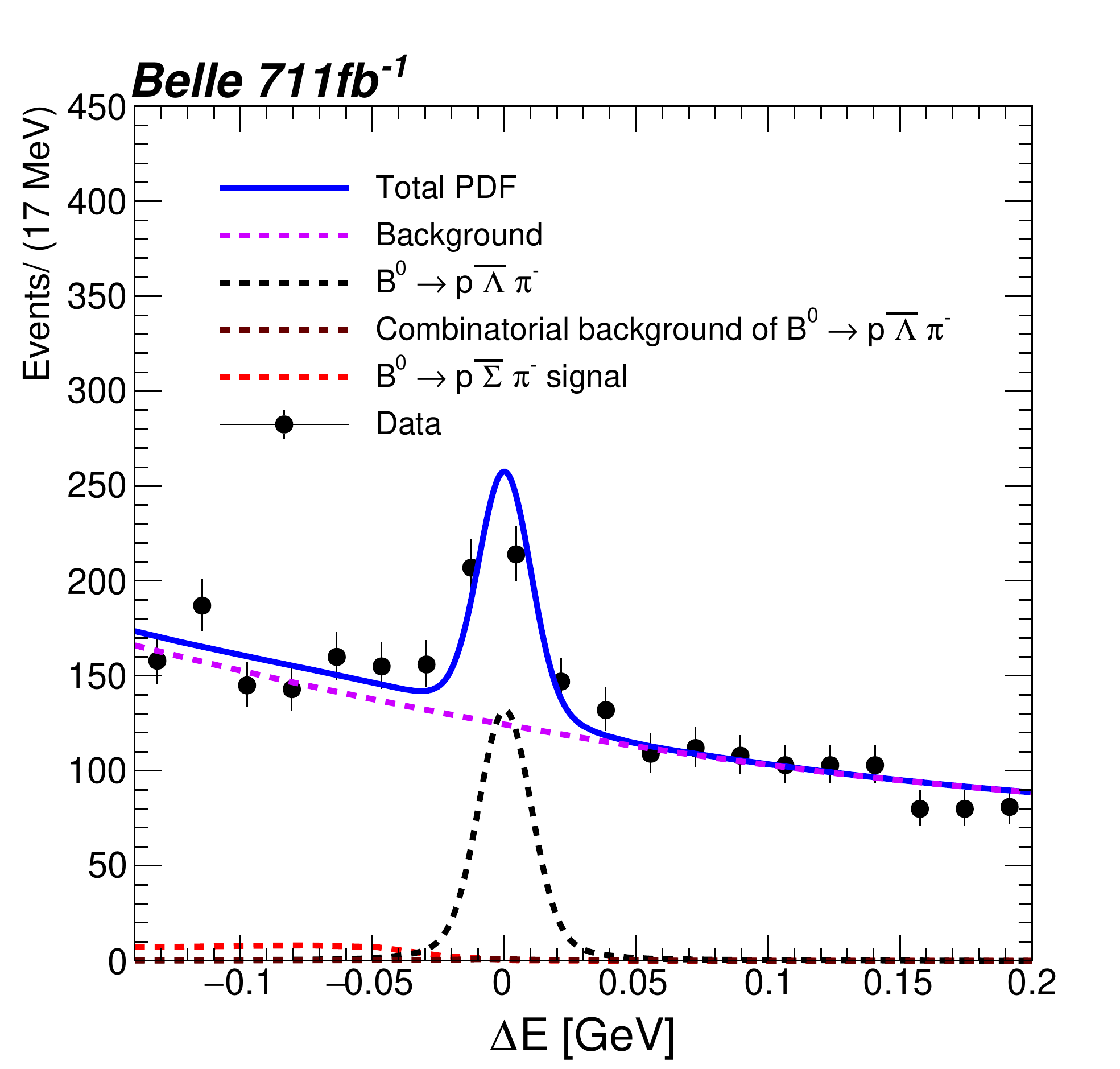}
\end{minipage}
}
\subfigure[]{
\begin{minipage}[c]{0.25\textwidth}%调节数值可以改变两图之间的距离
\centering
\includegraphics[width=4.5cm]{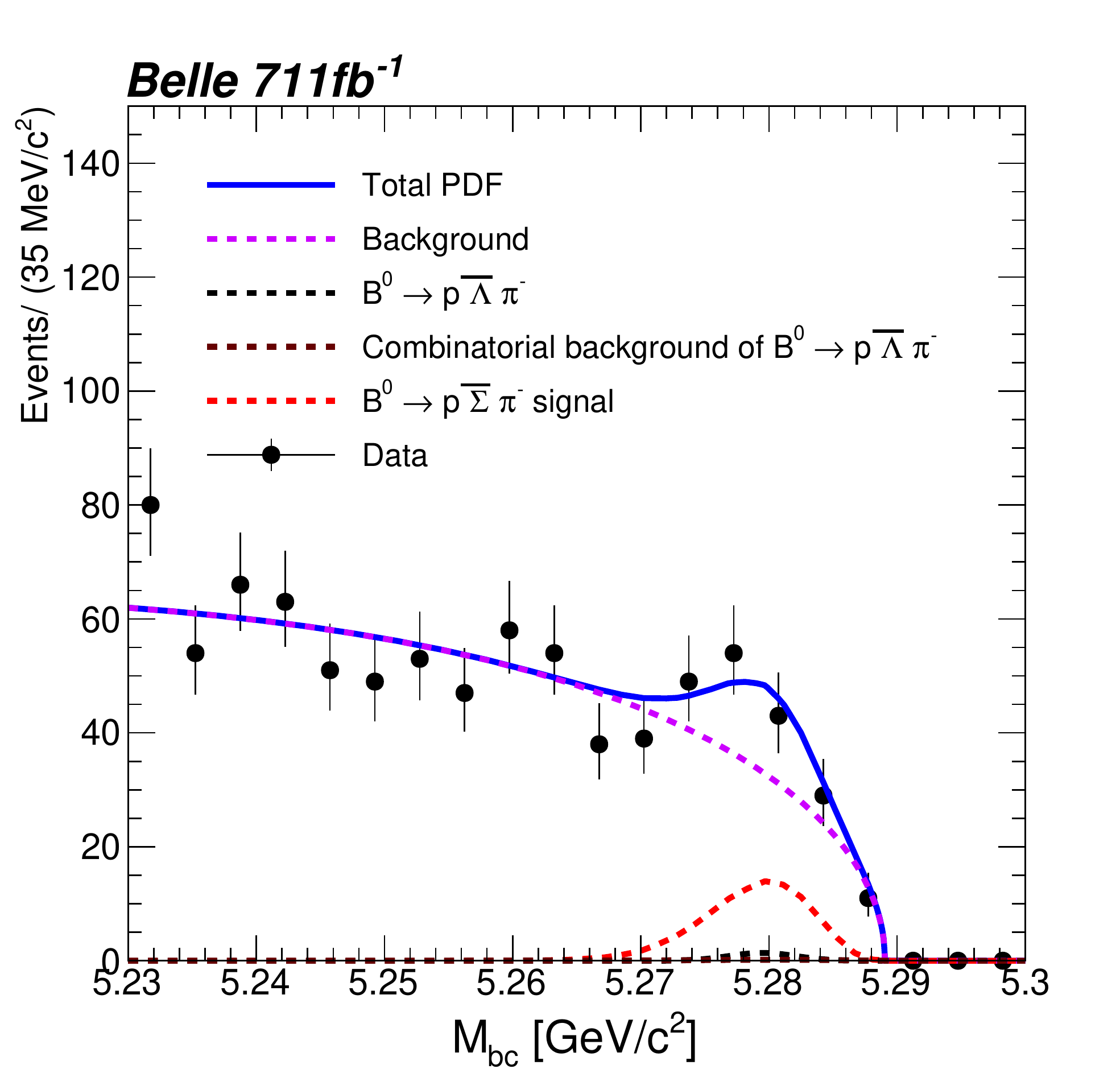}
\end{minipage}%
\begin{minipage}[c]{0.25\textwidth}
\centering
\includegraphics[width=4.5cm]{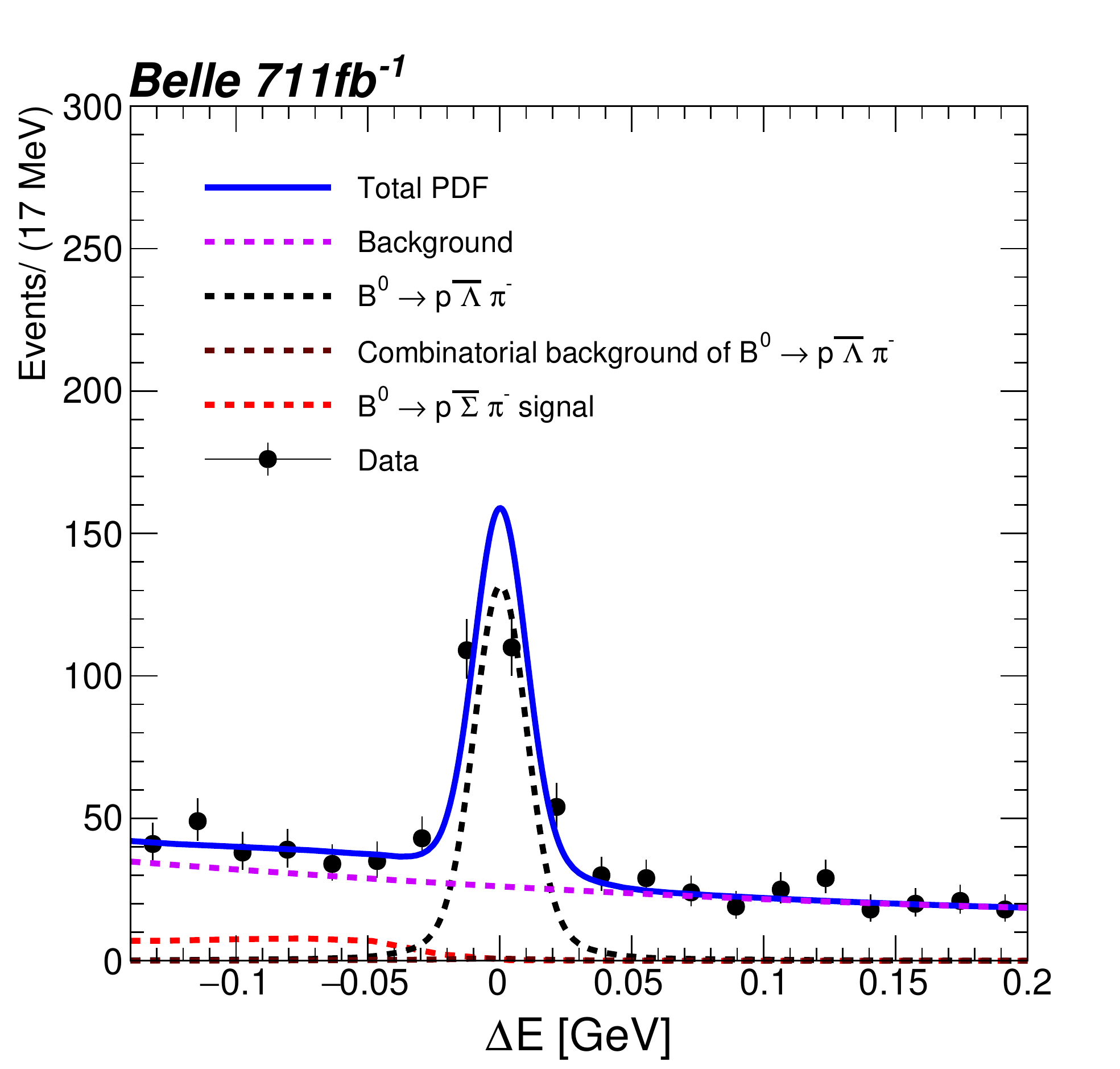}
\end{minipage}

}

\caption{ Projections of the fit results on $M_{\text{bc}}$ (left) and $\Delta E$ (right) distributions of the selected candidates (data points with error bars). Curves represent the components of the PDF in Eq~\ref{eq:2}: $B^0 \to p \bar{\Sigma}^0 \pi^-$ signal (red dashed line), combinatorial background from $B^0 \to p\bar{\Lambda} \pi^-$ (brown dashed line), $B^0 \to p\bar{\Lambda}\pi^-$ signal (black dashed line), background (purple dashed line), and total PDF (blue solid line). (a) The projected regions are the fitted regions: $\Delta E\in (-0.14,0.2)\ \text{GeV}$ for the left column and $M_\text{bc} \in (5.23,5.30)\ \text{GeV}/c^{2}$ for the right column. (b) The projected regions are $\Delta E\in (-0.14,-0.05)\ \text{GeV}$ for the left column and $M_\text{bc} \in (5.26,5.30)\ \text{GeV}/c^{2}$ for the right column to enhance the $B^0 \to p\bar{\Sigma}^0 \pi^-$ PDF component.}
\label{fig:result_whole}
\end{figure}

\begin{figure}[htb]
\centering
\subfigure[]
{
\begin{minipage}[c]{0.25\textwidth}%调节数值可以改变两图之间的距离
\centering
\includegraphics[width=4.5cm]{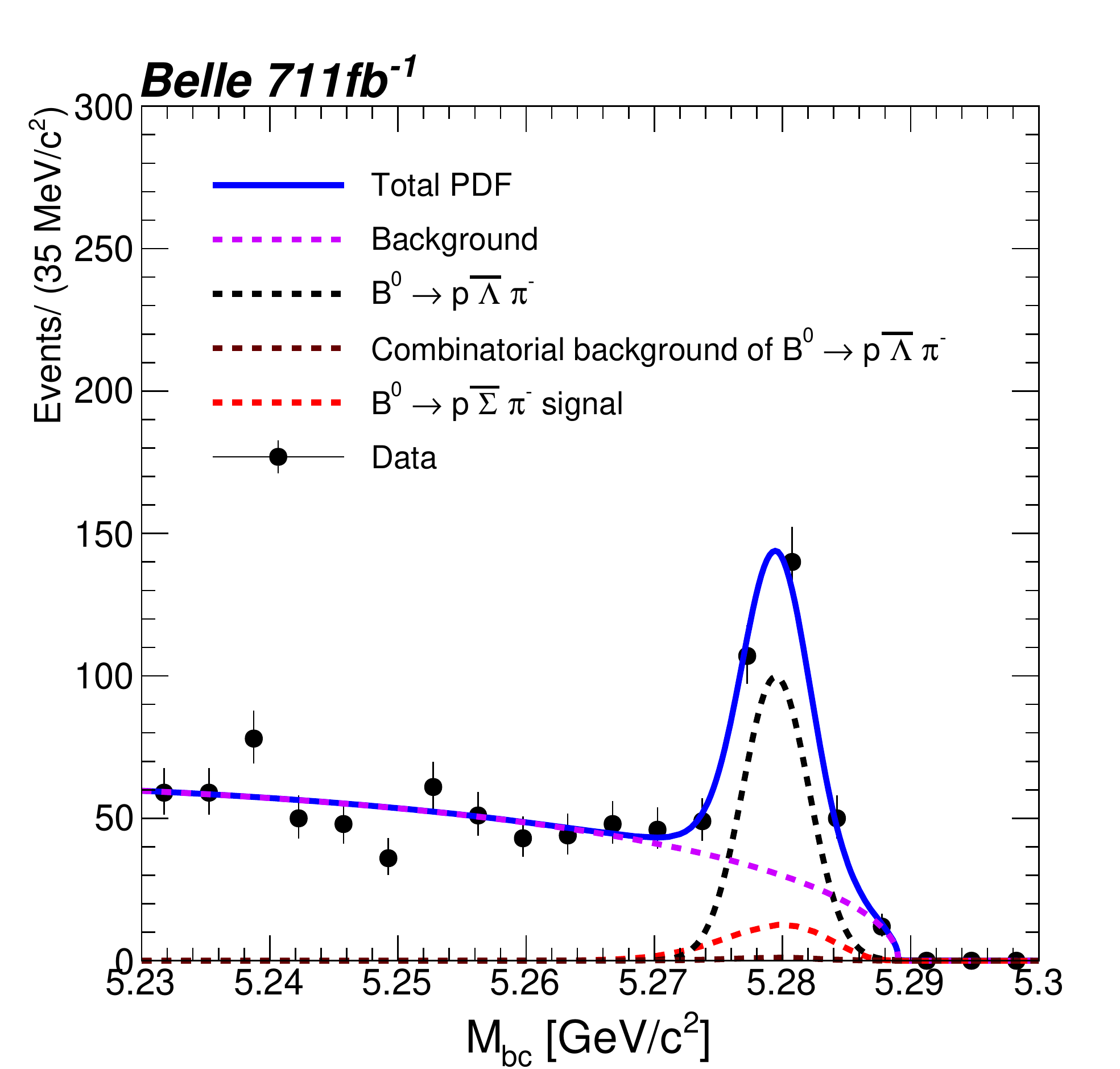}
\end{minipage}%
\begin{minipage}[c]{0.25\textwidth}
\centering
\includegraphics[width=4.5cm]{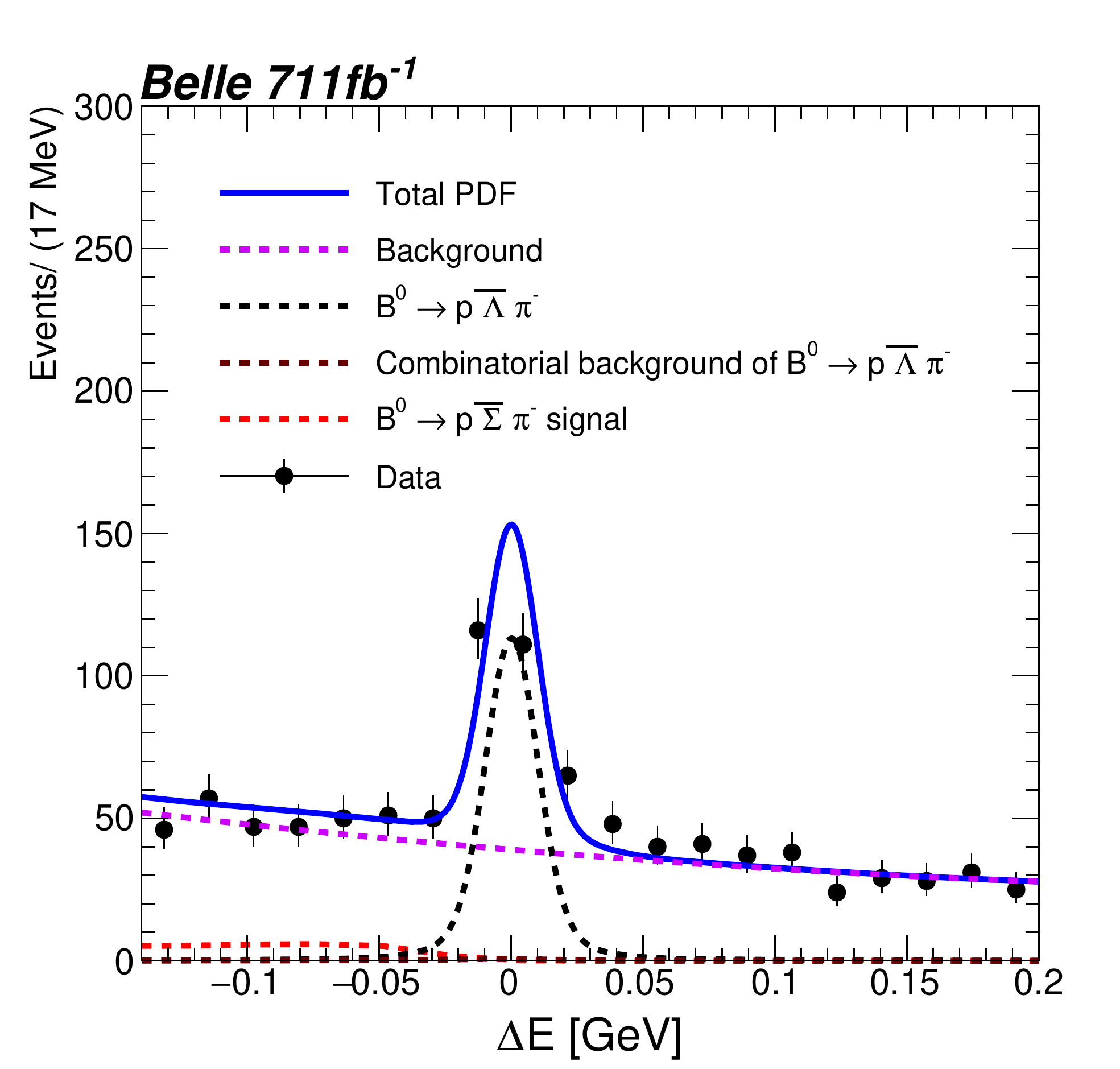}
\end{minipage}

}
\subfigure[]{
\begin{minipage}[c]{0.25\textwidth}%调节数值可以改变两图之间的距离
\centering
\includegraphics[width=4.5cm]{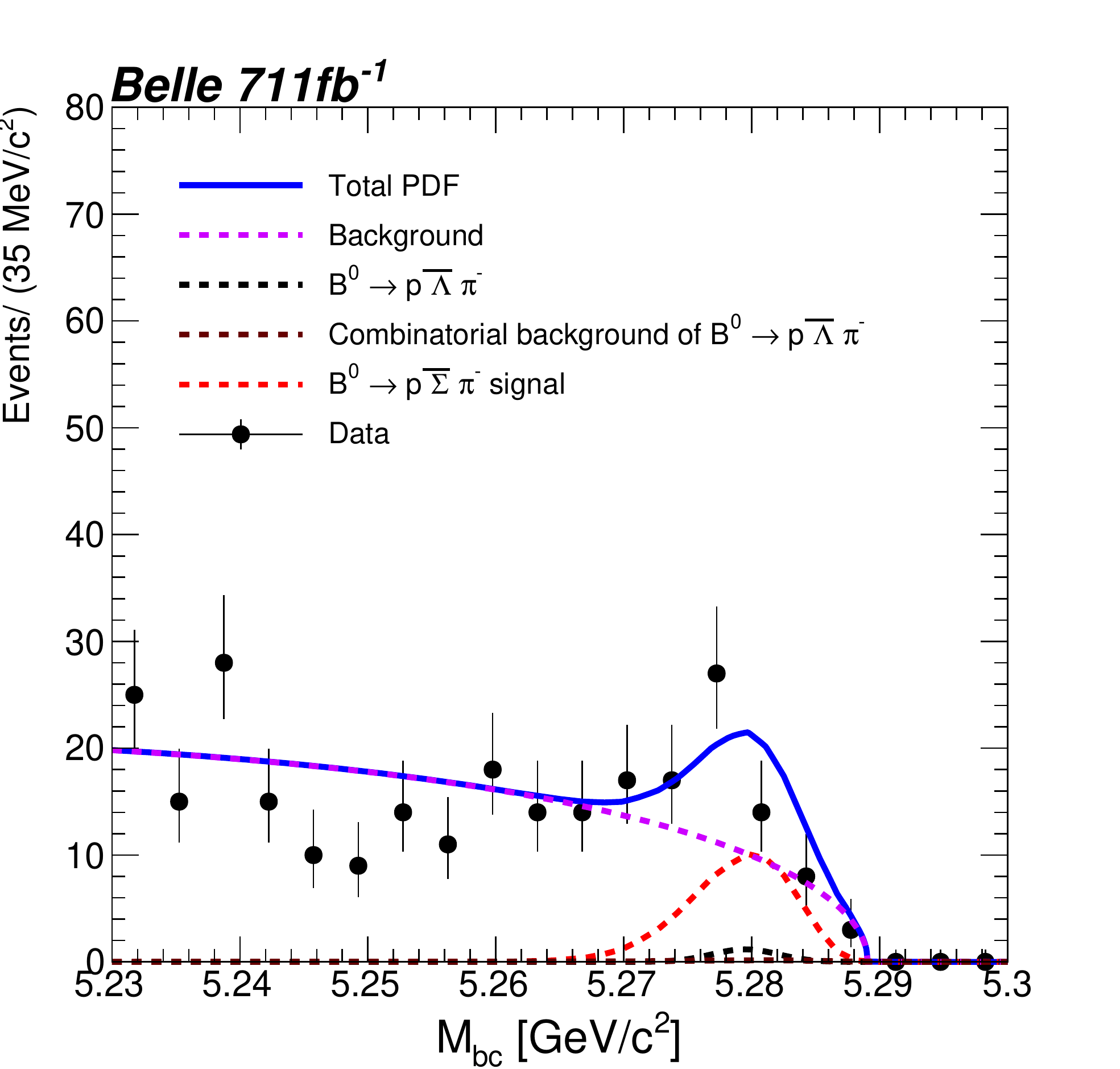}
\end{minipage}%
\begin{minipage}[c]{0.25\textwidth}
\centering
\includegraphics[width=4.5cm]{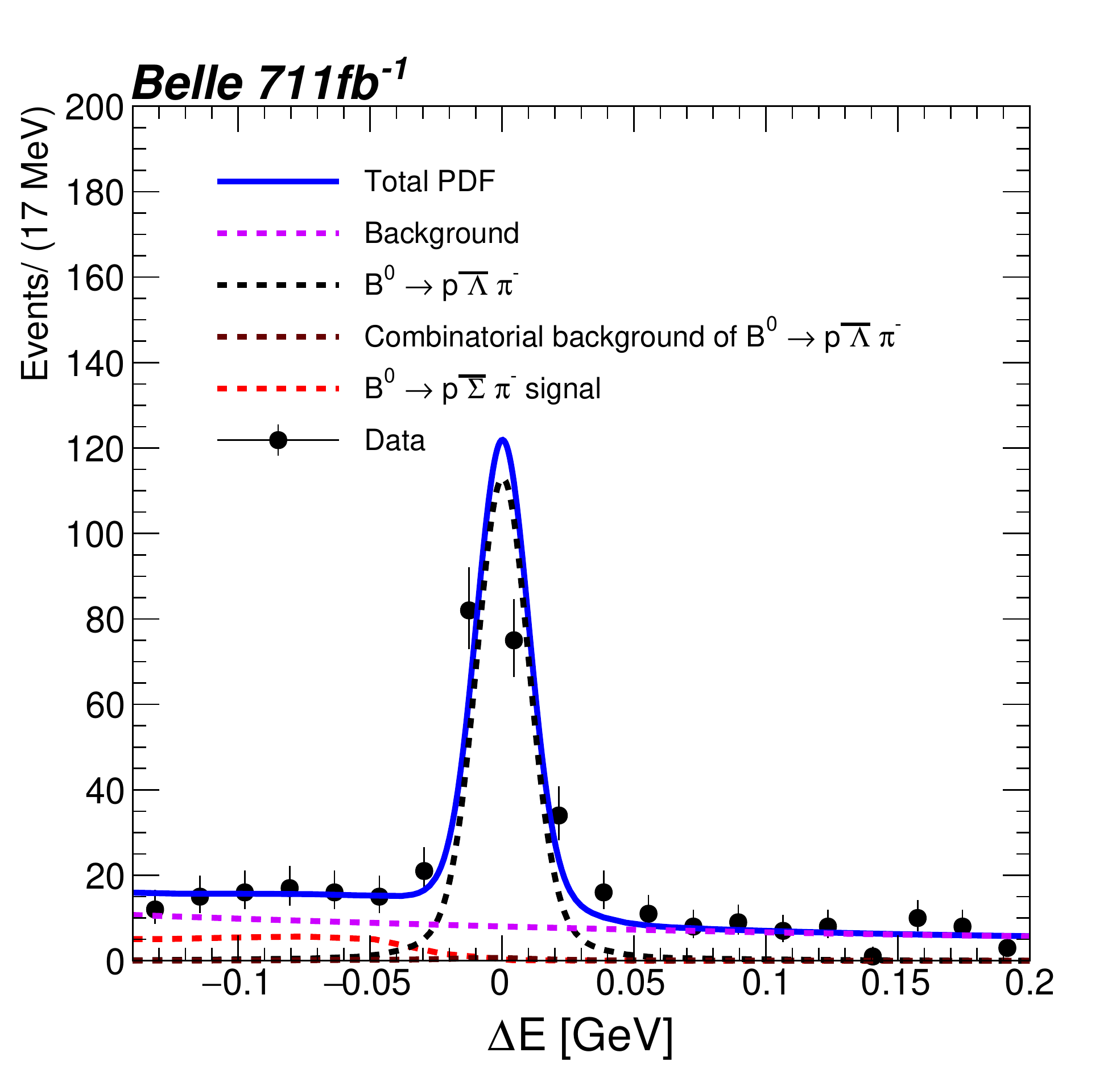}
\end{minipage}

}
\caption{The data points and curves shown are the same as those in
Figure 2, but with an additional selection $M_{p\bar{\Lambda}} < 2.8\ \text{GeV}/c^{2}$ applied. }
\label{fig:result_m_pl<2.8}
\end{figure}

Figure 2 shows projections of the fit of Eq.~\ref{eq:2} to data. We obtain signal yields of $N_{s1}=$ $50^{+18}_{-17}$ with a significance of $3.0\sigma$ including the relevant systematic uncertainty and $N_{s2}=$ $216\pm 17$. The value of the significance is defined by $\sqrt{-2 \ln[\mathcal{L}(\theta_{0})/\mathcal{L}({\hat{\theta}_{0})}]}$, where $\mathcal{L}(\theta_0)$ is the value of the likelihood function when the respective yield $\theta_{0}$ is set to 0 and $ \mathcal{L}(\hat\theta_0) $ when it is allowed to vary. In order to include the relevant systematic uncertainty in the significance, the values of the likelihood function in the significance calculation are smeared by using the uncertainty due to the PDF modeling described later. We also perform a fit in the TE region of  $M_{p\bar{\Lambda}}<2.8\ \text{GeV}/c^{2}$, which is shown in Figure \ref{fig:result_m_pl<2.8}. In this region, we obtain signal yields of $N_{s1}=$ $37^{+12}_{-11}$ with a significance of $3.5\sigma$ for $B^0\to p\bar{\Sigma}^0\pi^-$ and $N_{s2}=$ $185\pm15$ for $B^{0}\to p \bar{\Lambda}\pi^{-}$. The concentration of signal events in the TE region indicates that the TE effect also exists in $B^{0}\to p \bar{\Sigma}^0\pi^{-}$ decay. Due to the exclusion of soft photon in reconstruction of the $B^0\to p\bar{\Sigma}^0\pi^-$ channel, the $\Delta E$ distribution of that channel locates mostly in the negative $\Delta E$ region. We fit the regions of $\Delta E$ and $M_\text{bc}$ at $\Delta E\in (-0.14,-0.05)\ \text{GeV}$ and $M_\text{bc} \in (5.26,5.30)\ \text{GeV}/c^{2}$ to enhance the $B^0\to p\bar{\Sigma}^0\pi^-$ signal PDFs in Figure~\ref{fig:result_whole} and Figure~\ref{fig:result_m_pl<2.8}.

The branching fraction is calculated as 
\begin{equation} \label{eq:3}
    \mathcal{B} = \frac{ N_{\rm sig}}{N_{B\bar{B}}\ \epsilon\ \mathcal{C}_{\rm PID}\ 
    \mathcal{C}_{\rm bkg}},
\end{equation}
where $N_{\rm sig}$ is the number of signal events, $N_{B\bar{B}}$ = $772\times10^6$ is the number of $B\bar{B}$ pairs, $\epsilon$ denotes the signal efficiency obtained from signal MC samples, and $\mathcal{C}_{\rm PID}$, $\mathcal{C}_{\rm bkg}$ are efficiency calibration factors for the particle identification and the continuum suppression obtained from data, respectively. $\mathcal{C}_{\rm PID}$ is calculated as $\mathcal{C}_{p\rm ID} \mathcal{C}_{\pi\rm ID}$, where $\mathcal{C}_{p\rm ID}$ is the calibration factor for proton identification determined from a $\Lambda \to p\pi$ control sample, and $\mathcal{C}_{\pi\rm ID}$ is the calibration factor for charged-pion identification determined from a large $D^{*+} \to D^{0}(K^{-}\pi^{+})\pi^{+}$ control sample. $\mathcal{C}_{\rm bkg}$ is the calibration factor for the $q\bar{q}$ continuum background suppression using the neural-network classifier that is obtained using large $B^0\to D^{-}\pi^+$ and $D^{-}\to K^0_{S}\pi^-$ control samples. The calibration factors are summarized in Table~I. 

\begin{table}[htb]
    \centering
    \caption{Summary of efficiency calibration factors.}
    \begin{tabular}{c|c|c}
    \hline
    \hline
Source  & $B^0\to p\bar{\Lambda}\pi^-$ & $B^0\to p\bar{\Sigma}^0\pi^-$  \\
\hline
Proton ID  &    $0.95\pm0.01$   &  $0.94\pm 0.01$ \\ 
Pion ID    &    $0.94\pm0.01$   &  $0.94\pm 0.01$ \\
$q\bar{q}$ Continuum Suppression & $0.99\pm 0.02$   & $0.99\pm 0.02$ \\
\hline
Total Factor & $0.88\pm 0.02$ & $0.88\pm 0.02$ \\
    \hline
    \hline
    \end{tabular}
    \label{tab:my_label}
\end{table}

Different sources of systematic uncertainty are considered in this study. The independent uncertainties are summed in quadrature. The uncertainty on the number of $B\bar{B}$ pairs is $1.4\%$. The uncertainty from charged-particle tracking is 0.35$\%$ per track which is obtained from studying a sample of partially reconstructed $D^{*+}\to D^{0}(\pi^{-}\pi^{+}K^0_{S})\pi^{+}$ events. The uncertainty from proton identification is 0.6$\%$ for $B^0 \to p \bar{\Lambda} \pi^-$ and $B^0 \to p \bar{\Sigma}^0 \pi^-$. The uncertainty from pion identification is 0.7$\%$ for $B^0 \to p \bar{\Lambda} \pi^-$ and $B^0 \to p \bar{\Sigma}^0 \pi^-$. The uncertainty from the $\Lambda$ selection is 3.4$\%$, which is obtained from a control sample of  $B^{+}\to \Lambda\bar{\Lambda}K^+$ events. The uncertainty from the suppression of $q\bar{q}$ continuum background is 2.5$\%$, which is estimated using a control sample of $B^0\to D^{-}\pi^+(D^-\to K^0_{S}\pi^-)$ events, which have a final state that is topologically similar to $B^0 \to p \bar{\Lambda} \pi^-$. The uncertainty on the $\Lambda \to p \pi^-$ branching fraction is taken from the world average value \cite{Workman:2022ynf}.  The uncertainties due to the choice of model functions in the PDF in Eq~\ref{eq:2} are 1.4$\%$ for $B^0 \to p \bar{\Lambda} \pi^-$ and 3.3$\%$ for $B^0 \to p \bar{\Sigma}^0 \pi^-$ obtained by shifting the fixed parameters of the $B^0 \to p \bar{\Lambda}^0 \pi^-$ signal PDF $P_{s2}$ by 1$\sigma$, by changing the bin width of the histogram used to model the $B^0 \to p \bar{\Sigma}^0 \pi^-$ signal PDF $P_{s1}$, and by changing the background PDF $P_{b}$ from second-order polynomial to third-order polynomial.
The systematic uncertainties are summarized in Table \ref{tab:summary_sys}.\par

\begin{table}[ht]
    \centering
    \caption{Systematic uncertainties.}
    \begin{tabular}{c|c|c}
    \hline
    \hline
    Source & $B^0\to p\bar{\Lambda}\pi^-$ & $B^0\to p\bar{\Sigma}^0\pi^-$ \\
    \hline
     Number of $B\bar{B}$ & $1.4\%$  &  $1.4\%$ \\
     Tracking & $1.4\%$ & $1.4\%$ \\
     Proton ID  & $0.6\%$   &   $0.6\%$   \\ 
     Pion ID   &  $0.7\%$  &  $0.7\%$   \\ 
     $ \mathcal{B}(\Lambda \to p \pi^-$)& $0.8\%$  &   $0.8\%$   \\ 
     $\Lambda$ selection & $3.4\%$ & $3.4\%$ \\ 
     $q\bar{q}$ Continuum suppression &  $2.4\%$   & $2.4\%$ \\ 
     PDF modeling & 1.4$\%$ & 3.3$\%$ \\ 
     \hline
     Total uncertainty  & 5.0$\%$ & 5.8$\%$ \\
    \hline
    \hline
    \end{tabular}
    \label{tab:summary_sys}
\end{table}

\begin{figure}[htb]
    \centering
    \includegraphics[width =6cm]{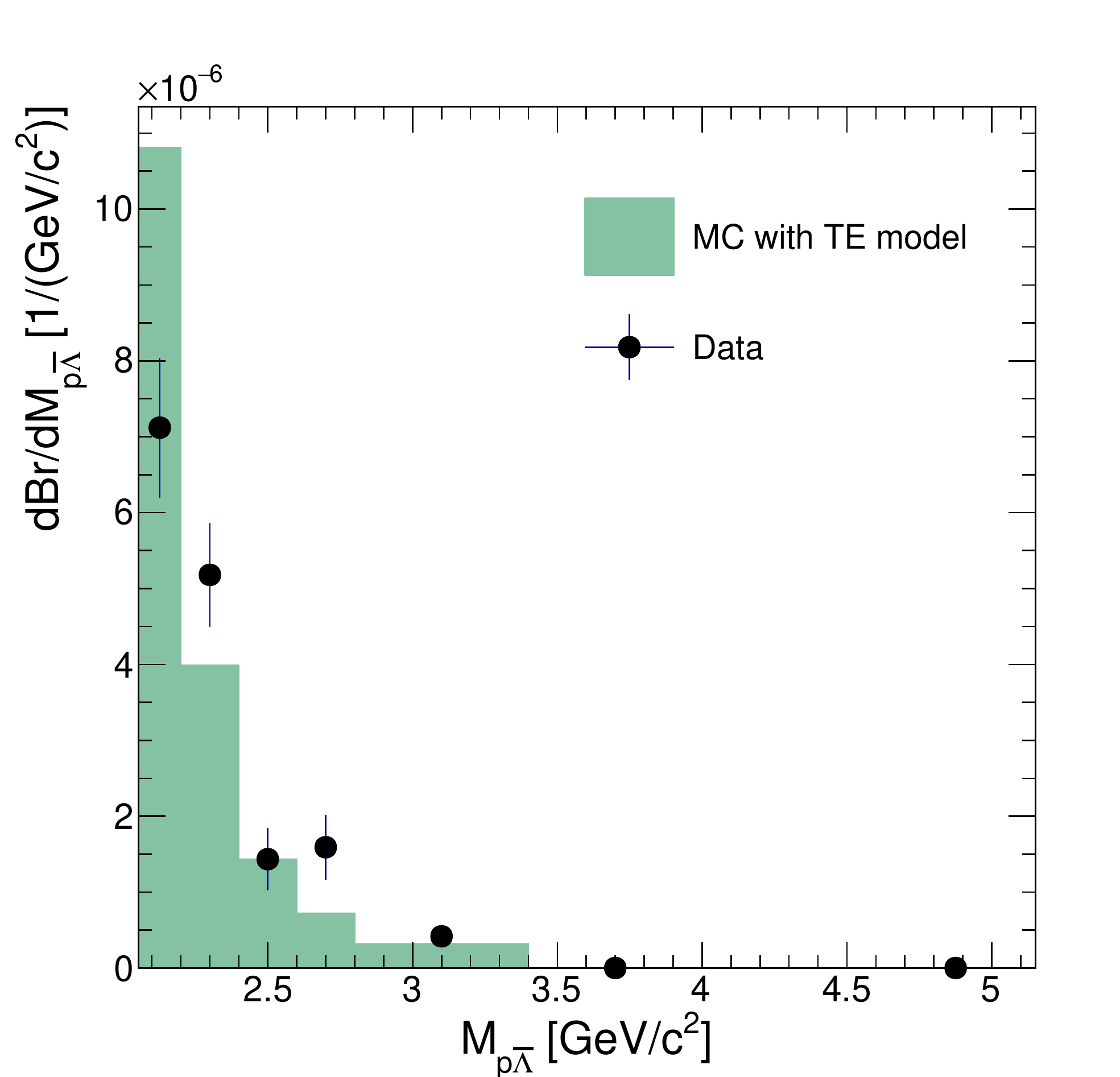}
    \caption{Differential branching fraction for $B^{0}\to p\bar{\Lambda}\pi^{-}$ as a function of $M_{p\bar{\Lambda}}$ (data points with statistical uncertainties). The shaded histogram represents the distribution obtained from the TE MC data sample. }
    \label{fig:m_pl}
\end{figure}

To study the $M_{p\Lambda}$ dependence of the efficiencies and the branching fractions, we fit Eq.~\ref{eq:2} in narrow $M_{p\Lambda}$ bins, and obtain the partial branching fractions based on Eq.~\ref{eq:3}. The results are listed in Table~\ref{tab:m_pl_plambdapi}. We sum these partial branching fractions in full $M_{p\bar{\Lambda}}$ region to obtain $\mathcal{B}( B^0 \to  p \bar\Lambda  \pi^- ) = (3.21^{+0.28}_{-0.25}(\text{stat})\pm 0.16(\text{syst})) \times 10^{-6}$ and $\mathcal{B}(B^0 \to p \bar\Sigma^0  \pi^-) = (1.17^{+0.43}_{-0.40}(\text{stat})\pm 0.07(\text{syst})) \times 10^{-6}$. In Fig. \ref{fig:m_pl}, we only show the $M_{p\bar{\Lambda}}$ dependence of $\mathcal{B}$($B^0\to p\bar{\Lambda}\pi^-$) because the yield of $B^{0}\to p\bar{\Sigma}^0\pi^{-}$ is insufficient to obtain precise differential branching fractions. \par

\begin{table}[htb]
    \centering
    \caption{Measured partial branching fractions $\mathcal{B}(10^{-6})$ in different $M_{p\bar{\Lambda}}$ bins.}
    \begin{tabular}{c|c|c}
    \hline
    \hline
   $M_{p\bar{\Lambda}}\ (\text{GeV}/c^{2})$ &$B^{0}\to p\bar{\Lambda}\pi^{-}$ & $B^0\to p\bar{\Sigma}^0\pi^-$ \\
    \hline
      $<$2.2  & $1.04^{+0.14}_{-0.13}$   & $0.27^{+0.12}_{-0.11}$  \\
     2.2--2.4   & $1.04^{+0.14}_{-0.13}$ & $0.24^{+0.14}_{-0.12}$    \\
     2.4--2.6   & $0.29^{+0.09}_{-0.07}$ & $0.13^{+0.13}_{-0.12}$    \\
     2.6--2.8   & $0.32^{+0.09}_{-0.08}$ & $0.12^{+0.13}_{-0.11}$    \\
     2.8--3.4   & $0.25^{+0.08}_{-0.07}$ & $0.29^{+0.19}_{-0.18}$    \\
     3.4--4.0   & $0.11^{+0.06}_{-0.05}$ & $-0.22^{+0.14}_{-0.12}$    \\
     4.0--4.6   & $0.07^{+0.06}_{-0.05}$ & $0.23^{+0.20}_{-0.17}$    \\
     $>$4.6     & $0.09^{+0.09}_{-0.07}$ & $0.11^{+0.20}_{-0.18}$          \\
     \hline
     $<$2.8     & $2.69^{+0.24}_{-0.21}$ &  $0.76^{+0.26}_{-0.23}$ \\
     Full region & $3.21^{+0.28}_{-0.25}$ & $1.17^{+0.43}_{-0.40}$ \\ 
    \hline
    \hline
    \end{tabular}
    \label{tab:m_pl_plambdapi}
\end{table}

\begin{figure}[htb]
    \centering
    \includegraphics[width = 6cm]{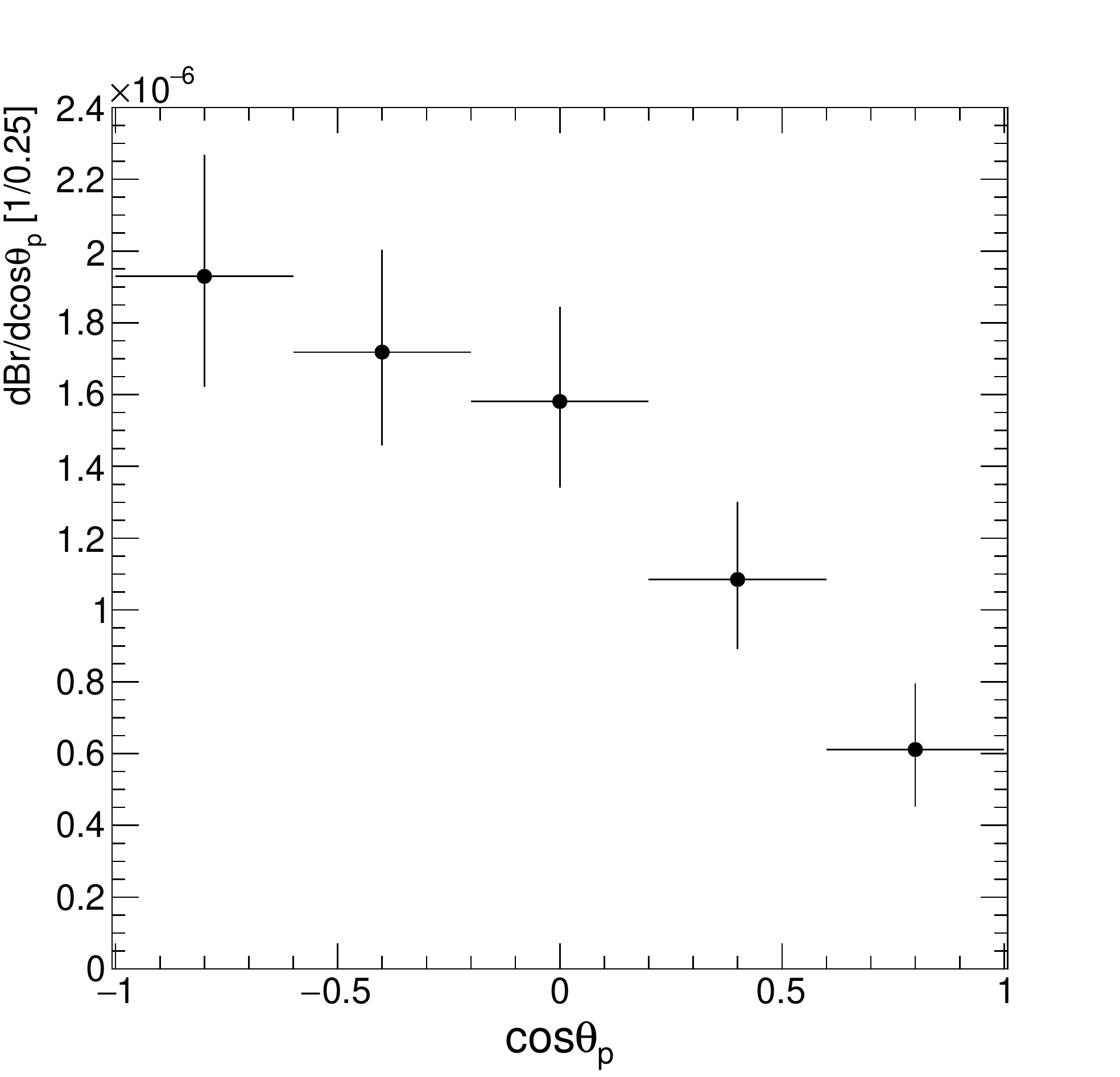}
    \caption{Differential branching fraction for $B^{0}\to p\bar{\Lambda}\pi^{-}$ as a function of $\cos\theta_{p}$ in the baryon-pair rest frame.}
    \label{fig:cos_p_asy}
\end{figure}

Using a similar approach, we study the branching fraction of $B^0 \to  p \bar\Lambda  \pi^-$ as a function of $\cos\theta_{p}$, where $\theta_{p}$ is the angle between the proton and the pion in the baryon pair rest frame. Table \ref{tab:cos_p_table} lists the partial branching fractions for $B^0\to p\bar{\Lambda}\pi^-$ and $B^0\to p\bar{\Sigma}^0\pi^-$ in each $\cos\theta_{p}$ bin. Figure \ref{fig:cos_p_asy} shows the $\cos\theta_{p}$ dependence of $\mathcal{B}( B^0 \to  p \bar\Lambda  \pi^- )$. Our measurement confirms the asymmetry of this distribution observed by the previous measurement \cite{PhysRevD.76.052004}. 

\begin{table}[htb]
    \centering
    \caption{Measured partial branching fractions $\mathcal{B}(10^{-6})$ in TE region and different $\cos\theta_{p}$ bins.}
    \begin{tabular}{c|c|c}
    \hline
    \hline
     $\cos\theta_{p}$ & $B^{0}\to p\bar{\Lambda}\pi^{-}$ &  $B^0\to p\bar{\Sigma}^0\pi^-$ \\
     \hline
      $-1.0$ to $0.6$ & $0.79^{+0.14}_{-0.13}$  & $0.20^{+0.13}_{-0.12}$  \\ 
      $-0.6$ to $0.2$ & $0.70^{+0.12}_{-0.11}$  & $0.23^{+0.11}_{-0.10}$  \\
      $-0.2$ to $0.2$ & $0.64^{+0.11}_{-0.10}$  & $0.09^{+0.11}_{-0.10}$  \\
      $0.2$ to $0.6$  & $0.44^{+0.09}_{-0.08}$  & $0.17^{+0.11}_{-0.10}$  \\
      $0.6$ to $1.0$  & $0.25^{+0.07}_{-0.06}$  & $0.05^{+0.11}_{-0.10}$  \\
    \hline
    \hline
    \end{tabular}
    \label{tab:cos_p_table}
\end{table}

In conclusion, we measure the branching fraction of $B^0\to p\bar{\Sigma}^0\pi^-$ for the first time to be $\mathcal{B}(B^0 \to p \bar \Sigma^0  \pi^-) = (1.17^{+0.43}_{-0.40}(\text{stat})\pm 0.07(\text{syst})) \times 10^{-6}$ with a significance of $3.0 \sigma$. For the yield in TE region, we obtain a slightly larger significance of 3.5$\sigma$. Both significances take into account systematic uncertainties. The $B^0\to p\bar{\Sigma}^0\pi^-$ signal yield is concentrated in the TE region indicating that a threshold enhancement is also present in the $B^0\to p\bar{\Sigma}^0\pi^-$ channel. For the $B^{0}\to p\bar{\Lambda}\pi^{-}$ channel, we measure a branching fraction of $(3.21^{+0.28}_{-0.25}(\text{stat}) \pm 0.16(\text{syst})) \times 10^{-6} $ which supersedes the previous measurement~\cite{PhysRevD.76.052004}, and the systematic uncertainty due to the normalized mode $B^{0}\to p\bar{\Lambda}\pi^{-}$ in LHCb rare baryonic decay measurement could be improved by $20\%$. We also study the differential branching fraction as a function of $M_{p\bar{\Lambda}}$ and $\cos\theta_{p}$, and observe an asymmetric distribution in $\cos\theta_{p}$. Our study also shows that $B^0\to p\bar{\Sigma}^0\pi^-$ has a lower branching fraction than $B^{0}\to p\bar{\Lambda}\pi^{-}$ which is in contrast to the QCD model prediction \cite{PhysRevD.66.014020,PhysRevD.66.054004}, but more close to the modified approach~\cite{Hou}. 

This work, based on data collected using the Belle detector, which was
operated until June 2010, was supported by 
the Ministry of Education, Culture, Sports, Science, and
Technology (MEXT) of Japan, the Japan Society for the 
Promotion of Science (JSPS), and the Tau-Lepton Physics 
Research Center of Nagoya University; 
the Australian Research Council including grants
DP210101900, % Urquijo
DP210102831, % Sevior
DE220100462, % Hsu
LE210100098, % Infrastructure
LE230100085; % Infrastructure
Austrian Federal Ministry of Education, Science and Research (FWF) and
FWF Austrian Science Fund No.~P~31361-N36;
National Key R\&D Program of China under Contract No.~2022YFA1601903,
National Natural Science Foundation of China and research grants
No.~11575017,
No.~11761141009, 
No.~11705209, 
No.~11975076, 
No.~12135005, 
No.~12150004, 
No.~12161141008, 
and
No.~12175041, 
and Shandong Provincial Natural Science Foundation Project ZR2022JQ02;
the Ministry of Education, Youth and Sports of the Czech
Republic under Contract No.~LTT17020;
the Czech Science Foundation Grant No. 22-18469S;
Horizon 2020 ERC Advanced Grant No.~884719 and ERC Starting Grant No.~947006 ``InterLeptons'' (European Union);
the Carl Zeiss Foundation, the Deutsche Forschungsgemeinschaft, the
Excellence Cluster Universe, and the VolkswagenStiftung;
the Department of Atomic Energy (Project Identification No. RTI 4002) and the Department of Science and Technology of India; 
the Istituto Nazionale di Fisica Nucleare of Italy; 
National Research Foundation (NRF) of Korea Grant
Nos.~2016R1\-D1A1B\-02012900, 2018R1\-A2B\-3003643,
2018R1\-A6A1A\-06024970, RS\-2022\-00197659,
2019R1\-I1A3A\-01058933, 2021R1\-A6A1A\-03043957,
2021R1\-F1A\-1060423, 2021R1\-F1A\-1064008, 2022R1\-A2C\-1003993;
Radiation Science Research Institute, Foreign Large-size Research Facility Application Supporting project, the Global Science Experimental Data Hub Center of the Korea Institute of Science and Technology Information and KREONET/GLORIAD;
the Polish Ministry of Science and Higher Education and 
the National Science Center;
the Ministry of Science and Higher Education of the Russian Federation, Agreement 14.W03.31.0026, % from 15.02.2018
and the HSE University Basic Research Program, Moscow; % from 15.04.2021
University of Tabuk research grants
S-1440-0321, S-0256-1438, and S-0280-1439 (Saudi Arabia);
the Slovenian Research Agency Grant Nos. J1-9124 and P1-0135;
Ikerbasque, Basque Foundation for Science, Spain;
the Swiss National Science Foundation; 
the Ministry of Education and the Ministry of Science and Technology of Taiwan;
and the United States Department of Energy and the National Science Foundation.
These acknowledgements are not to be interpreted as an endorsement of any
statement made by any of our institutes, funding agencies, governments, or
their representatives.
We thank the KEKB group for the excellent operation of the
accelerator; the KEK cryogenics group for the efficient
operation of the solenoid; and the KEK computer group and the Pacific Northwest National
Laboratory (PNNL) Environmental Molecular Sciences Laboratory (EMSL)
computing group for strong computing support; and the National
Institute of Informatics, and Science Information NETwork 6 (SINET6) for
valuable network support.

%***** Acknowledgments *****

%\bibliographystyle{unsrt}
\bibliography{ref}

\end{document}